\documentclass[12pt,osajnl2,preprint,showpacs,superscriptaddress]{revtex4}  
\usepackage[draft]{hyperref} 

\begin{document}
\newcommand{\be}{\begin{equation}}
\newcommand{\ee}{\end{equation}}
\newcommand{\wt}{\widetilde}
\newcommand{\ba}[1]{\left(\begin{array}{#1}}
\newcommand{\ea}{\end{array}\right)}

\title{\bf A POVM view of the ensemble approach to polarization 
optics}

\author{Sudha}
\affiliation{Department of P.G. Studies in Physics, Kuvempu University, 
Shankaraghatta-577 451, India} 
\email{arss@rediffmail.com}
\author{ A.V. Gopala Rao}
\affiliation{Department of Studies in Physics, Manasagangothri, 
University of Mysore, Mysore 570 006, India}

\author{ A. R. Usha Devi} 
\affiliation{ Department of Physics, Jnanabharathi Campus,  Bangalore University, 
Bangalore-560 056, India}
\author{A.K. Rajagopal}
\affiliation{ Department of Computer Science and Center for Quantum
Studies,  George Mason University, Fairfax, VA 22030, USA \\ and
Inspire Institute Inc., McLean, VA 22101, USA}

\begin{abstract}

Statistical ensemble formalism  of Kim, Mandel and Wolf (J. Opt. Soc. Am. A {\bf 4}, 433 (1987))  
offers
a realistic model for  characterizing the effect of stochastic non-image forming 
optical media on the state of polarization of  transmitted light. 
With suitable choice of the Jones ensemble, various  Mueller transformations - 
some of which have been unknown so far -  are deduced. 
It is observed that the ensemble approach is formally identical to 
 the positive operator valued measures (POVM) on the
 quantum density matrix.
 This observation, in combination with the recent suggestion by Ahnert and Payne 
 (\pra {\bf 71}, 012330, (2005)) - in the context of generalized quantum measurement on single 
photon polarization states -
  that linear optics elements can be employed in setting up all possible POVMs, enables us 
 to propose a way of realizing different types of Mueller devices.
 
\end{abstract}
\maketitle

\section{Introduction}
The intensity and polarization of a beam of light passing through
an isolated optical device undergoes a linear transformation.
But this is an ideal situation because, in
general, the optical system is embedded in some media such as
atmosphere or other ambient material, which further modifies the polarization 
properties of the light beam  passing through it. A
statistical ensemble model describing random linear optical media was
formulated two decades ago by Kim, Mandel and Wolf~\cite{kim}, but is not
examined in any detail in the literature, to the best of our
knowledge. The purpose of the present paper is to pursue this avenue in a 
new way arising from the realization of a relationship, presented here, with the 
positive operator valued measures (POVM) of quantum measurement theory. 
This is because the transformation of the polarization states 
of a light beam propagating through an 
ensemble of deterministic  optical devices exhibits 
a structural similarity with the POVM
transformation of quantum density matrices. 
This connection motivates, in view of the recent interest in the implementations of POVMs on
single photon density matrix employing linear optics elements~\cite{povm},  identification of  
experimental schemes to realize various kinds of Muller transformations. 
 The properties of the transformation of the 
  polarization states of light form a much studied topic in literature~[3 -- 17]. 
  Thus the power of the ensemble approach becomes evident 
  in elucidating the known optical devices as well as some hitherto unknown types~\cite{avg}, 
  which had remained only 
  a mathematical possibility.

The contents of this paper are organized as follows. 
In Sec.~2,  a concise formulation of the Jones and Mueller matrix theory, along with a summary of  
main 
results of Gopala Rao et al.~\cite{avg} is given. Based on the approach of  Kim, Mandel and 
Wolf~\cite{kim} 
 suitable Jones ensembles,  corresponding  to various types  of  Mueller transformations are 
identified in Sec.~3. 
In Sec.~4, a structural  equivalence between Jones ensemble and 
  POVMs of quantum measurement theory is established. Following the linear optics 
 scheme of Ahnert and Payne~\cite{povm} for the implementation of POVMs on single photon density 
matrix,  
 experimental setup for realizing  Mueller matrices of types I and II are suggested in Sec.~5. 
 The final section has some concluding remarks.

\section{Brief summary of known results on the Jones and the Mueller formalism.}
Following the standard procedure, let $E_1$ and
$E_2$,  defined here as a column matrix $ {\bf E}=\ba{l} E_1 \\
E_2 \ea$,  denote  two components of the transverse electric field
vector associated with a light beam. The coherency matrix (or the polarization matrix) of the 
light beam  is a positive semidefinite 2x2 hermitian matrix defined by, 
\be
\label{e1}
 {\bf C}=\langle {\bf E}\otimes
{\bf E}^\dagger\rangle.
 \ee 

Expressing this in terms of the standard Pauli matrices $\sigma_1=\ba{ll}0 & 1 \\ 1 & 0 \ea,$  
 $\sigma_2~=~\ba{cc}0 & -i \\ i & 0 \ea, \   \sigma_3=\ba{cc}1 &
0
\\ 0 & -1 \ea $ and the unit matrix $ \sigma_0=\ba{ll}1 & 0
\\ 0 & 1 \ea$, we have 
\be
\label{e2}
{\bf C}=\frac{1}{2} \sum_{i=0}^3 s_i{\bf
\sigma}_i=\frac{1}{2}\ba{cc} s_0+s_3 & s_1-is_2 \\
s_1+is_2 & s_0-s_3 \ea 
\ee
The physical significance of the quantities arising here are 
\be
\label{e3}
\begin{array}{l}
s_0={\rm Tr}\;({\bf C \sigma_0})={\rm Intensity\  of\  the\  beam } \\
s_i={\rm Tr}\; ({\bf C\sigma_i})={\rm Components\  of\
Polarization\ vector\ {\vec{\bf s}}\  of\  the\
 beam}
 \end{array}
 \ee
Thus the coherency matrix completely specifies the physical
properties of the light beam. The four-vector ${\bf S}=\ba{c} s_0
\\ {\vec{\bf s}}\ea$ defined by Eq. (\ref{e3}) is the well known
Stokes vector, which represents the state of polarization of the light beam. Because $\bf C$ is 
hermitian, the Stokes vector is
real. The positive semidefiniteness of $\bf C$ implies that the Stokes
vector must satisfy the properties 
\be
\label{e4} 
s_0 > 0, \;
s_0^2-|{\vec{\bf
s}}|^2\geq 0 \ee

A 2x2 complex matrix $\bf J$, called the Jones matrix, represents
the so-called deterministic optical device~\cite{azzam} or medium. 
When a light beam represented by $\bf E$ passes through such a
medium, the transformed light beam is given by ${\bf E}'={\bf
JE}$. Correspondingly, the coherency matrix $\bf C$ transforms as 
\be
\label{e5}
{\bf C}'={\bf J}{\bf C}{\bf J}^\dagger \ee
(Here ${\bf J}^\dagger$ is the hermitian conjugate of $\bf J$.)

Alternatively, instead of the $2\times 2$ matrix transformation of
the coherency matrix,  as given by  Eq. (5), 
a transformation 
\be
\label{e6} {\bf S}'={\bf MS} \ee
of the four componental Stokes column ${\bf S}$  
through a real 4x4 matrix $\bf M$, called the Mueller
matrix, is found be more useful~\cite{azzam}.

Using Eq. (\ref{e3}) and
Eq. (\ref{e5} we have,
\[ s'_i={\rm Tr}({\bf C'}{\bf \sigma}_i)={\mbox Tr}({\bf JCJ}^\dagger\,{\bf 
\sigma}_i)=\frac{1}{2}\sum_{j=0}^3 {\rm Tr}({\bf J}^\dagger{\bf \sigma_i}{\bf J}{\bf \sigma}_j)s_j
\]
which leads to the well-known relationship \cite{kim}
\[ M_{ij}=\frac{1}{2}\,{\rm Tr}({\bf J}^\dagger{\bf \sigma_i}{\bf
J}{\bf \sigma}_j)\] between the elements of a Jones matrix and that of 
corresponding  Mueller matrix.

But in the case where medium cannot be represented by a Jones
matrix, it is not possible to characterize the change in the state of polarization 
of the light beam through Eq. (\ref{e5}). In such a situation, Mueller formalism provides a 
general approach  for  the polarization 
transformation of the light beam. The Mueller matrix $\bf M$  
is said to be  non-deterministic when it has no corresponding Jones characterization. 

Mathematically, a Mueller device can be represented by 
 any $4\times 4$ matrix such that the Stokes parameters of the outgoing light beam satisfy 
 the physical constraint  Eq. (\ref{e4}). 
 In other words, a Mueller matrix is any $4\times 4$ real matrix that transforms a
Stokes vector into another Stokes vector.
There are many aspects of the relationships between these two  
formulations of the polarization optics and a complete
characterization of Mueller matrices has been the subject matter
of Ref.~[1, 3-17]. It was Gopala Rao et al.~\cite{avg} who presented 
a complete set of necessary and sufficient conditions for any 4x4 real matrix to be a
Mueller matrix. In so doing, they found that there are two
algebraic types of Mueller matrices called type I and type II; 
and it has been shown~\cite{avg} that only a subset of the type-I Mueller matrices - called 
deterministic or pure Mueller matrices -  have corresponding Jones characterization.    
All the known polarizing optical devices such as retarders, polarizers, analyzers, optical 
rotators are pure Mueller type   and  are well understood. 
Mueller matrices of the Type II variety  are yet to be physically realized and have remained
as mere mathematical possibility. For the sake of completeness, we
present here the characterization as well as categorization of these
two types of Mueller matrices as is given in Ref.~\cite{avg}. This will enable us 
to  show  that both Type I and II  Mueller devices are realizable  
 in an unified manner in terms of the proposed ensemble
approach~\cite{kim}.

\noindent {\bf I. A $4\times 4$ real matrix $\bf M$ is called a 
type-I
Mueller matrix iff}
\begin{description}
\item{(i)} $M_{00} \geq 0$ \item{(ii)} The G-eigenvalues
$\rho_0,\;\rho_1,\;\rho_2,\;\rho_3$ of the  matrix
$\bf N$=${\widetilde{\bf M}}{\bf GM}$ are all real. (Here, 
$\widetilde {\bf M}$ stands for the transpose of $\bf M$; 
G-eigenvalues are the eigenvalues of the matrix ${\bf GN}$, with ${\bf G}={\rm diag}(1,\, -1,\, 
-1,\, -1)$).
\item{iii)} The largest G-eigenvalue $\rho_0$ possesses a
time-like G-eigenvector and the G-eigenspace of $\bf N$ contains
one time-like and three space-like G-eigenvectors.
\end{description}
\noindent {\bf II. A $4\times 4$ real matrix ${\bf M}$ is called a
type-II Mueller matrix iff}
\begin{description}
\item{(i)} $M_{00} > 0.$
\item{(ii)} The G-eigenvalues $\rho_0,\;\rho_1,\;\rho_2,\;\rho_3$ 
 of  $\bf N$=${\widetilde{\bf M}}{\bf GM}$
are all real. 
\item {(iii)} The largest G-eigenvalue $\rho_0$
possesses a null G-eigenvector and the G-eigenspace of $\bf N$
contains one null and two space-like G-eigenvectors. 
\item{(iv)} If ${\bf X}_0={\bf e}_0+{\bf e}_1$ is the null G-eigenvector of $\bf N$ such that 
${\bf e}_0$ is a time-like vector with positive zeroth component, ${\bf e}_1$ is a space-like 
vector G-orthogonal to ${\bf e}_0$ 
then  ${\widetilde{\bf e}_0}{\bf N}{\bf e}_0 >0$.  
\end{description}

Despite the knowledge of these new category of Mueller matrices~\cite{cvm,avg},  
not much attention is paid for realizing the corresponding  devices. An
experimental arrangement involving a parallel combination of
deterministic (pure Mueller) optical devices  is proposed
in Ref.~\cite{avg} for realizing type-II Mueller devices.  
The physical situations, where the beam of light is subjected to 
the influence of a medium such as atmosphere was addressed in   
Ref.~\cite{kim}. In the next section, we discuss this ensemble approach  
for random optical media, 
proposed by Kim, Mandel and Wolf~\cite{kim} .

\section{Mueller matrices as ensemble of Jones devices}

Kim et. al.~\cite{kim} associate a set of probabilities $\{p_e,\;\sum p_e=1 \}$ 
to describe the stochastic medium. Then a Jones device 
${\bf J}_e$ associated with each element $e$ of the ensemble gives a
corresponding coherency matrix ${\bf C'}_e={\bf J}_e{\bf C}{\bf
J}_e^\dagger$. The ensemble averaged coherency matrix 
\be 
\label{e7}
{\bf C}_{av}=\sum_e p_e {\bf C'}_e=\sum_e p_e
({\bf J}_e{\bf C}{\bf J}_e^\dagger) \ee 
then describes the effects of the medium on the beam of light. In
a similar fashion, the corresponding ensemble of Mueller matrices
$\{{\bf M}_e\}$ associated with the ensemble of Jones matrices
$\{{\bf J}_e\}$ is constructed and its ensemble averaged Mueller
matrix is similarly formed as ${\bf M}_{av}=\sum_e p_e{\bf
M}_e$. Since a linear combination of Mueller matrices with
non-negative coefficients is also a Mueller matrix, the ensemble
averaged Mueller matrix ${\bf M}_{av}$ is a Mueller matrix~\footnote{This is
because, each Mueller matrix ${\bf M}_e$ transforms an initial Stokes
vector  into a final Stokes vector and a linear combination of
Stokes vectors with non-negative coefficients $p_e$ is again a Stokes
vector.}.

We now turn to the question of constructing an appropriate
ensemble designed to describe a given physical situation. The
simplest example of an ensemble is one where the elements 
 are chosen entirely randomly, i.e., the system is
described by a chaotic ensemble where the probabilities are all
equal, $p_e=\frac{1}{n},$ where $n$ denotes the number of elements in the ensemble. 
The coherency matrix ${\bf C}_{av}$ of
the light beam passing through  such a chaotic assembly is just an
arithmetic average of the coherency matrices ${\bf C'}_e={\bf
J}_e{\bf C}{\bf J}_e^\dagger$ and hence 
\be 
\label{e8}
{\bf C}_{av}={\frac{1}{n}}\sum_{e=1}^{n}{\bf J}_e{\bf C}{\bf
J}_e^\dagger
\ee
More general models can be constructed depending on the
medium for the propagation of the beam of light. For example,
one may employ various types of filters or solid state systems
through which the light passes; the assignment of the
Jones matrices and the corresponding probabilities will then
differ depending on the weights placed on these
elements.

Restricting ourselves to an ensemble consisting of only two Jones
devices which occur with equal probability $p_1=1/2$, $p_2=1/2$,
we have found out that the resultant Mueller matrices can either
be deterministic  or non-deterministic. 
  We give in the foregoing (see Table I) some examples of 
 Mueller matrices corresponding to different choices of Jones matrices in an ensemble
{${\bf J}_e, \ e=1,2,$} for some representative cases. This will also serve to
show the generality of the ensemble procedure in capturing the
physical realizations for the Mueller devices discussed in Ref.~\cite{avg}.

\begin{table}[htb]
\centering\caption{Mueller matrices resulting from
2-element Jones ensemble.}
{\scriptsize
\begin{tabular}{ccccc}
\hline 
  & ${\bf J}_1$ & ${\bf J}_2$ & ${\bf M}=p_1{\bf
M}_1+p_2{\bf M}_2,$ & Type of ${\bf M}$ \\
& & &  $p_1=p_2=\frac{1}{2}.$ & \\
\hline 
1. & $\frac{1}{\sqrt 6}\ba{cc} 1 & 1-i  \\
                                  1+i & -1  \ea, $ & $\frac{1}{\sqrt 6}\ba{cc} 1 & 1-i  \\
                                  1+i & -1  \ea $ & $ \frac{1}{3}\ba {rrrr}  3 & 0 & 0 & 0 \\
                                  0 & -1 & 2 & 2  \\
                                  0 & 2 & -1 & 2 \\
                                  0 & 2 & 2 & -1 \ea $ & Pure Mueller
                                  \\ \hline
2. & $\ba{cc} 1 & 0  \\
                                  0 & 0  \ea $ & $ \ba{cc} 0 & 1  \\
                                  0 & 0  \ea $ &  $\frac{1}{2}\ba {rrrr} 1 & 0 & 0 & 0 \\
                                  1 & 0 & 0 & 0  \\
                                  0 & 0 & 0 & 0 \\
                                  0 & 0 & 0 & 0 \ea $ & Type-I
                                  \\ \hline

3. & $\frac{1}{\sqrt 2}\ba{cc} 0 & 1  \\
                                  1 & 0  \ea $ & $\frac{1}{\sqrt 2} \ba{cc} 1 & 0  \\
                                  0 & -1  \ea $ &  $\frac{1}{2}\ba {rrrr} 1 & 0 & 0 & 0 \\
                                  0 & 0 & 0 & 0  \\
                                  0 & 0 & 0 & 0 \\
                                  0 & 0 & 0 & -1 \ea $ & Type-I
                                  \\ \hline
4. & $\frac{1}{\sqrt 3}\ba{cc} 0 & 1  \\
                                  1 & 0  \ea $ & $\frac{1}{\sqrt 3}\ba{cc} 1 & -i  \\
                                  i & -1  \ea $ &  $ \frac{1}{6}\ba {rrrr}  3 & 0 & 0 & 0 \\
                                  0 & -1 & 0 & 2  \\
                                  0 & 0 & -1 & 0 \\
                                  0 & 2 & 0 & -1 \ea $ & Type-I
                                  \\ \hline

5. & $\frac{1}{\sqrt 5}\ba{cc} 1 & 1-i  \\
                                  1+i & -1  \ea $ & $\frac{1}{\sqrt 5}\ba{cc} 1 & -i  \\
                                  i & -1  \ea $ &  $ \frac{1}{10}\ba {rrrr}  5 & 0 & 0 & 0 \\
                                  0 & -1 & 2 & 4  \\
                                  0 & 2 & -3 & 2 \\
                                  0 & 4 & 2 & -1 \ea $ & Type-I
                                  \\ \hline

6. & $\ba{cc} 0 & 1  \\
                                  0 & 0  \ea $ & $\ba{cc} 0 & 0  \\
                                  0 & 1  \ea $ & $ \frac{1}{2} \ba {rrrr}  1 & -1 & 0 & 0 \\
                                  0 & 0 & 0 & 0  \\
                                  0 & 0 & 0 & 0 \\
                                  0 & 0 & 0 & 0 \ea $ & Type-II
                                  \\ \hline
7. & $\frac{1}{2}\ba{cc} 1 & 1  \\
                                  1 & -1  \ea $ & $\frac{1}{2}\ba{cc} 1 & -i  \\
                                  i & -1  \ea $ &  $ \frac{1}{4}\ba {rrrr}  2 & 0 & 0 & 1 \\
                                  0 & 0 & 1 & 0  \\
                                  0 & 1 & 0 & 0 \\
                                  1 & 0 & 0 & 0 \ea $ & Type-II
                                  \\ \hline

\end{tabular}}
\end{table}

In Table I, the Jones matrices chosen are so as to give pure
Mueller (deterministic) and non-deterministic type-I, type-II matrices 
respectively. We observe that an assembly of Jones matrices can result in
a pure Mueller matrix if and only if all elements of the assembly
correspond to the same optical device. This is because, with all
${\bf J}_e$'s are same,  a transformation of the form ${\bf C}_{av}=\sum_e p_e ({\bf J}_e
{\bf C} {\bf J}_e^\dagger)$ is equivalent to a transformation of
the Stokes vector $\bf S$ through a Mueller matrix ${\bf
M}_{av}=\sum p_e {\bf M}_e={\bf M}_{\rm pure}$. 
When the medium is represented by a pure
Mueller matrix, the outgoing light beam will have the same degree
of polarization as the incoming light beam. In fact, pure Mueller
matrix is the simplest among type-I Mueller matrices. Not all
type-I Mueller matrices preserve the degree of polarization of the
incident light beam. To see this, note that  the type-I matrix of
example 2 (see Table I) converts any incident light beam into 
a linearly polarized light beam; the other three type-I matrices 
(examples 3 to 5)  transform  completely polarized
light beams into partially polarized light beams. Similarly,
type-II Mueller matrices do not, in general, preserve the degree of 
polarization of the incident light beam. It may be seen that 
the type-II Mueller matrix of example 7 is a depolarizer matrix, since 
it converts any incident light beam into an unpolarized light beam. 

Though one cannot a priori state which choices of Jones matrices
result in type-I or type-II, it is interesting to observe that
all types of Mueller matrices result - even in 2-element ensembles.
It is not difficult to conclude that an ensemble, with more Jones
devices and  with different weight factors, can give rise
to a variety of Mueller matrices of all possible
algebraic types. It would certainly be interesting to physically
realize such systems.

In the following section, a connection between the ensemble approach for optical devices 
and the POVMs of quantum measurement theory is established.

\section{A connection to Positive Operator Valued Measures}

We will now show that the phenomenology of the ensemble
construction of Kim, Mandel and Wolf~\cite{kim} described above has a
fundamental theoretical underpinning, if we make a formal
identification of the coherency matrix with the density matrix
description of the subsystem of a composite quantum system. The
coherency matrix defined by Eqs. (\ref{e1}) and (\ref{e2}) 
resembles a quantum density
matrix in that both describe a physical system by a hermitian,
trace-class, and positive semi-definite matrix. While the 
 quantum density matrix has unit trace, the coherency
matrix has intensity of the beam as the value of the trace. The
Jones matrix transformation is a general transformation of the
coherency matrix, which preserves its hermiticity and positive
semi-definiteness - but changes the values of the elements of the
coherency matrix. The most general transformation of the
density matrix ${\bf \rho}$, which preserves its hermiticity,
positive semi-definiteness and also the unit trace is the
positive operator valued measures (POVM)~\cite{nc}: 
\be
\label{e9} 
{\bf \rho}'=\sum_{i=1}^n {\bf V}_{i}{\bf \rho}{\bf V}_i^\dagger;\ \ \ \ \ 
\sum_{i=1}^n {\bf V}_i^\dagger{\bf V}_i={\bf I} \ee
where $V_i$'s are general matrices and $\bf I$ is the unit element
in the Hilbert space. More generally, one could relax the condition of preservation of
the unit trace of the density matrix by examining the possibility
of a contracting transformation, where the unit matrix condition on
the POVM operators is replaced by an inequality.
 
This mathematical theorem has a physical basis in the 
Kraus operator formalism~\cite{nc} when we
consider the Hamiltonian description of a composite interacting
system $A,\ B$ described by a density matrix ${\bf \rho}(A,\;B)$ and deduce the
subsystem density matrix of  $A$ given by, ${\bf \rho}(A)={\rm
Tr}_{B}\;{\bf \rho}(A,\;B)$. In this case, the Kraus operators are
the explicit expressions of the POVM operators and contain the
effects of interaction between the systems $A$ and $B$ in the description of the 
subsystem $A$. It is thus clear that the
phenomenology of Ref.~\cite{kim} has a correspondence with
the Kraus formulation and the POVM theory. In order to make this
association complete, we compare Eq. (\ref{e9}) with the expression given
by Eq. (\ref{e7}). Apart from a phase factor, the Kraus operators $\{{\bf V}_i\}$,  
 associated with POVMs,   may be related to the 
 Jones assembly $\{{\bf J}_i\}$, chosen in the form  
 \be
 \label{e10}
{\bf V}_i={\sqrt p_i} {\bf J}_i, \ \ \ \  \sum_{i=1}^n {\bf
V}_i^\dagger{\bf V}_i=\sum_{i=1}^{n}p_i {\bf J}_i^\dagger{\bf J}_i
\ee
In the construction of the Table I presented  earlier, a simple
model was proposed where all probabilities were chosen to be equal
and the condition on the sum over the Jones matrix combinations
was set equal to unit matrix. In such cases, the intensity of
the beam  gets reduced by 1/n and the  polarization properties of the beam gets changed  as was 
described earlier. With this identification, we have provided here an important interpretation
and meaning to the phenomenology of the ensemble approach of Kim
et al.\cite{kim}.

Recently Ahnert and Payne \cite{povm} proposed  an experimental scheme to implement all possible 
POVMs on 
 single photon polarization states using linear optical elements. In view of  the connection 
between the 
ensemble formalism for  Jones and Mueller matrices with the POVMs,  a possible experimental 
realization 
of the two types of Mueller matrices is suggested  in the next section.

\section{Possible experimental realization of types I and II Mueller matrices.  }
We first observe that the density matrix of a single photon polarization state,  
\be 
\label{e11}
{\bf \rho}=\rho_{HH}\vert H \rangle\, \langle H\vert + \rho_{HV}\vert H \rangle\, \langle V\vert
+\rho^*_{HV}\vert V \rangle\, \langle H\vert +\rho_{VV}
\vert V \rangle\, \langle V\vert 
 \ee
is nothing but the coherency matrix of the photon~\cite{MW} 
\be
\label{e12}
{\bf C}=\ba{cc}  \langle {\hat {\bf a}^{\dagger}_H}\hat {\bf a}_H\rangle 
& \langle {\hat {\bf a}^{\dagger}_H}\hat {\bf a}_V\rangle \\
\langle {\hat {\bf a}^{\dagger}_V}\hat {\bf a}_H\rangle  & 
\langle {\hat {\bf a}^{\dagger}_V}\hat {\bf a}_V\rangle \ea ,
\ee
 where $\hat{\bf a}_H$ and $\hat{\bf a}_V$ are the creation operators of the polarization states  
of the single photon; $\{\vert H \rangle,\, \vert V \rangle\}$ denote the transverse orthogonal 
polarization states of photon. 
This is seen explicitly by noting that the average values of the Stokes operators are 
obtained as,  
\begin{eqnarray}
\label{e13}
s_0=\langle \hat {\bf S}_0 \rangle&=&\langle( {\hat {\bf a}^{\dagger}_H}\hat {\bf a}_H+ 
{\hat {\bf a}^{\dagger}_V}\hat {\bf a}_V )\rangle=\rho_{HH}+\rho_{VV}={\rm Tr}({\bf \rho}), 
\nonumber \\
s_1=\langle \hat {\bf S}_1 \rangle&=&\langle ({\hat {\bf a}^{\dagger}_H}\hat {\bf a}_V+ 
{\hat {\bf a}^{\dagger}_V}\hat {\bf a}_H )\rangle=\rho_{HV}+\rho^*_{HV}= 
{\rm Tr}({\bf \rho}\, \sigma_1),
\nonumber \\
s_2=\langle \hat {\bf S}_2 \rangle&=&i\, \langle ({\hat {\bf a}^{\dagger}_V}\hat {\bf a}_H- 
{\hat {\bf a}^{\dagger}_H}\hat {\bf a}_V )\rangle=i\, (\rho_{HV}-\rho^*_{HV})
={\rm Tr}({\bf \rho}\, \sigma_2),
\nonumber \\
s_3=\langle \hat {\bf S}_3 \rangle&=&\langle ({\hat {\bf a}^{\dagger}_H}\hat {\bf a}_H
- {\hat {\bf a}^{\dagger}_V}\hat {\bf a}_V )\rangle=\rho_{HH}-\rho_{VV}=
{\rm Tr}({\bf \rho}\, \sigma_3). 
\end{eqnarray}
Hence the proposed setup \cite{povm}, involving 
only linear optics elements such as polarizing beam splitters, 
rotators and phase shifters, that promises to implement all possible POVMs on a single photon 
polarization state leads to 
all possible ensemble realizations for the Mueller matrices. More specifically, this provides a 
general experimental scheme to realize varieties of Mueller matrices -  including the hitherto 
unreported type-II Mueller matrices. 
We briefly describe the  scheme proposed in Ref.~\cite{povm} and illustrate, by way of examples, 
how it leads to both type-I and type-II Mueller matrices. 

In Ref.~\cite{povm}, a module corresponds to an arrangement having  polarization beam splitters,  
polarization rotators,  phase shifters and unitary operators. For an $n$ element POVM, a setup 
involving $n-1$ 
modules are needed. That means, a single module is enough for a 2 element POVM; a setup involving 
two modules is required for a 3 element POVM and so on. 
We describe two, three element POVMs by specifying the optical elements in the 
respective modules and by specifying the corresponding
Kraus operators in terms of these elements. 

For any two operator POVM, the Kraus operators ${\bf V}_1$, ${\bf V}_2$ are given by ${\bf 
V}_1={\bf U}'{\bf D}_1{\bf U}$ and 
${\bf V}_2={\bf U}''{\bf D}_2{\bf U}$. Here ${\bf U}$, ${\bf U}'$, ${\bf U}''$ are the three 
 unitary operators in a single module. Denoting  $\theta$, $\phi$ as the angles of  rotation of 
the two 
variable polarization rotators and $\gamma$, $\xi$, the 
angles of the two variable phase shifters in the module, 
the diagonal matrices ${\bf D}_1$, ${\bf D}_2$ are given by,  
\be
\label{e14}
{\bf D}_1=\ba{cc} e^{i\gamma} \cos \theta & 0 \\ 
                 0 & \cos \phi \ea  , \ \           {\bf D}_2=\ba{cc} e^{i\xi} \sin \theta & 0 \\ 
                                                               0 & \sin \phi \ea
															      \ee 
The POVM elements 
\be 
\label{e15}
{\bf F}_1={\bf V}_1^\dagger{\bf V}_1={\bf U}^{\dagger}{\bf D}_1^{\dagger}{\bf D}_1{\bf U}, \ \  
{\bf F}_2={\bf V}_2^\dagger{\bf V}_2={\bf U}^{\dagger}
{\bf D}_2^{\dagger}{\bf D}_2{\bf U}
\ee 
satisfy the condition $\sum_{i=1,2}\, {\bf F}_i={\bf F}_1+{\bf F}_2={\bf I}$. 

For any three operator POVM, the Kraus operators are given by 
\be 
\label{e16}
\begin{array}{l}
{\bf V}_1={\bf U}'_{\rm I}{\bf D}_{\rm I}{\bf U}_{\rm I}, \\
{\bf V}_2={\bf U}'_{\rm II}{\bf D}_{\rm II}{\bf U}_{II}{\bf U}''_{\rm I}{\bf D}'_{\rm I}{\bf 
U}_{\rm I}, \\
{\bf V}_3={\bf U}''_{\rm II}{\bf D}'_{\rm II}{\bf U}_{\rm II}{\bf U}''_{\rm I}{\bf D}'_{\rm I}{\bf 
U}_{\rm I}, 
\end{array}  
\ee 
Here, the diagonal $\bf D$ matrices are 
\be 
\label{e17}
{\bf D}_{\rm I}=\ba{cc} e^{{i\gamma}_{\rm I}} \cos \theta_{\rm I}  & 0 \\ 
                 0 & \cos \phi_{\rm I} \ea  , \ \           {\bf D}'_{\rm I}=\ba{cc} 
e^{{i\xi}_{\rm I}} \sin \theta_{\rm I} & 0 \\ 
                                                               0 & \sin \phi_{\rm I} \ea 
\ee 
and
\be 
\label{e18}
{\bf D}_{\rm II}=\ba{cc} e^{{i\gamma}_{\rm II}} \cos \theta_{\rm II}  & 0 \\ 
                 0 & \cos \phi_{\rm II} \ea  , \ \           {\bf D}'_{\rm II}=\ba{cc} 
e^{{i\xi}_{\rm II}} \sin \theta_{\rm II} & 0 \\ 
                                                               0 & \sin \phi_{\rm II} \ea 
\ee 
($\theta_{\rm I}$, $\phi_{\rm I}$), ($\gamma_{\rm I}$, $\xi_{\rm I}$) are  respectively  the pair 
of  angles corresponding to variable polarization 
rotators and variable phase shifters in the first module. Similarly, ($\theta_{\rm II}$, 
$\phi_{\rm II}$), ($\gamma_{\rm II}$, $\xi_{\rm II}$) are  
the pairs of angles corresponding to variable polarization rotators and variable phase shifters 
respectively in the second module. ${\bf U}_{\rm I}$, 
${\bf U}'_{\rm I}$, ${\bf U}''_{\rm I}$ are the unitary operators used in the first module and  
${\bf U}_{\rm II}$, ${\bf U}'_{\rm II}$, ${\bf U}''_{\rm II}$ 
are the unitary operators used in the second module. (Notice that all the unitary operators in the 
above schemes are arbitrary and  a particular choice  
of the associated unitary operators gives rise to a  
different experimental arrangement). 
The extension of this scheme to n operator POVM involving n-1 modules is quite similar and is 
given in \cite{povm}. 

We had identified, in Sec.~3, that an ensemble average of 
Jones devices will lead to all possible types of Mueller matrices, some examples of which are 
given in Table 1. We now show that the experimental set up proposed   
in Ref.~\cite{povm} can also be used to realize varieties of Mueller devices. To substantiate our 
claim, 
we identify here the linear optical elements needed in the single module 
set up of Ahnert and Payne~\cite{povm}, which lead to the physical 
realization of two typical Mueller matrices given in Table 1. 

To obtain the type-I Mueller matrix  $\bf M$= $\frac{1}{2}\ba{rrrr} 1 & 0 & 0 & 0 \\
                                  0 & 0 & 0 & 0  \\
                                  0 & 0 & 0 & 0 \\
                                  0 & 0 & 0 & -1 \ea $ of example 3 (see  Table I), 
we use ${\bf U}={\bf I}$, ${\bf U}'=\ba{cc} 0 & 1 \\                                                                                                    
1 & 0 \ea$ and ${\bf U}''=\ba{cr} 1 & 0 \\                                                                                                             
0 & -1 \ea$ as the required unitary Jones devices and 
both the variable polarization rotators are set with their rotation angles 
$\theta$=$\phi=\pi/4$. There is no need of phase shifter 
devices in this case i.e,  $\gamma$=$\xi$=0. 

Similarly for the type-II Mueller matrix  $\bf M$= $\frac{1}{2}\ba{rrrr} 2 & 0 & 0 & 1 \\
                                  0 & 0 & 1 & 0  \\
                                  0 & 1 & 0 & 0 \\
                                  1 & 0 & 0 & 0 \ea $  of example 7, 
we find that ${\bf U}={\bf I}$, ${\bf U}'=\frac{1}{\sqrt 2} \ba{cr}
1 & 1 \\                                                                                                             
1 & -1 \ea$ and
 ${\bf U}''=\frac{1}{\sqrt 2} \ba{cc} 1 & i\\ 
i & 1 \ea$ are the required unitary Jones devices.  
The rotation angles of the variable polarization rotators are, as in the earlier case,  
$\theta$=$\phi=\pi/4$ and  there is no need of 
phase shifter devices i.e., $\gamma$=$\xi$=0. Notice that in both the above examples the unitary 
operators ${\bf U'},\ {\bf U''}$ correspond to linear and circular retarders~\cite{azzam}.

These two examples illustrate that  the experimental set up given in  Ref.~\cite{povm}  
may be utilized to realize the required non-determinisitc Mueller devices.   
In fact Mueller matrices corresponding to an ensemble with more than two Jones 
devices may also be realized  by employing  larger number of  modules as given in 
the experimental scheme proposed by Ref.~\cite{povm}.

\section{Conclusion}
We have established here a connection between the
phenomenological ensemble approach~\cite{kim} for the coherency matrix and the
POVM transformation of quantum density matrix. This opens up a fresh avenue
 to physically realize   types I and II of the
Mueller matrix classification of Ref.~\cite{avg}. We have also given experimental 
setup to implement Mueller transformations corresponding to ensemble average of 
Jones devices by employing the POVM scheme on the single photon density matrix suggested in 
Ref.~\cite{povm}, in the context of quantum measurement theory. 
It is gratifying to note that two decades after the introduction of the 
ensemble approach, which had remained obscure and only received passing 
reference in textbooks such as~\cite{MW}, its value is revealed in this 
paper through its connection with the new developments in quantum 
measurement theory.   We plan on exploring further the POVM transformation 
in the description of quantum polarization optics.

\end{document}